\documentclass[12pt,preprint]{aastex}
\usepackage{emulateapj5}
\usepackage{apjfonts}
\usepackage{epsfig}
\usepackage{natbib}

\newcommand{\chisq}{\ensuremath{\chi^2}}

\newcommand{\etal}{et al.}

\newcommand{\delsig}{$\Delta \sigma$}

\def\gtrsim{\mathrel{\hbox{\rlap{\hbox{\lower4pt\hbox{$\sim$}}}\hbox{\raise2pt\hbox{$>$}}}}}
\newcommand{\halpha}{H\ensuremath{\alpha}}
\newcommand{\hii}{\ion{H}{2}}
\newcommand{\hbeta}{H\ensuremath{\beta}}

\newcommand{\kms}{km s\ensuremath{^{-1}}}

\newcommand{\lf}{\ensuremath{L_{5100}}}

\newcommand{\ledd}{\ensuremath{L\mathrm{_{Edd}}}}
\newcommand{\lledd}{\ensuremath{L_{\mathrm{bol}}/L{\mathrm{_{Edd}}}}}
\newcommand{\lbol}{\ensuremath{L_{\mathrm{bol}}}}
\newcommand{\loiii}{\ensuremath{L_{\mathrm{[O {\tiny III}]}}}}
\newcommand{\lha}{\ensuremath{L_{\mathrm{H{\alpha}}}}}

\newcommand{\mbh}{\ensuremath{M_\mathrm{BH}}}
\newcommand{\meansig}{\ensuremath{\langle \sigma_{\rm
g}/\sigma_{\ast} \rangle}}

\newcommand{\msigma}{\ensuremath{M_{\mathrm{BH}}-\sigmastar}}
\newcommand{\msun}{\ensuremath{M_{\odot}}}

\newcommand{\oii}{[\ion{O}{2}]}
\newcommand{\oiii}{[\ion{O}{3}]}

\newcommand{\sii}{[\ion{S}{2}]}
\newcommand{\sigmastar}{\ensuremath{\sigma_{\ast}}}
\newcommand{\sigmagas}{\ensuremath{\sigma_{\rm g}}}

\newcommand{\vc}{$v_\mathrm{c}$}

\def\lax{{$\mathrel{\hbox{\rlap{\hbox{\lower4pt\hbox{$\sim$}}}\hbox{$<$}}}$}}
\def\gax{{$\mathrel{\hbox{\rlap{\hbox{\lower4pt\hbox{$\sim$}}}\hbox{$>$}}}$}}

\slugcomment{to appear in {\it The Astrophysical Journal}.}
\lefthead{Greene \& Ho}
\righthead{Gas and Stellar Kinematics in AGNs}
\shorttitle{Gas and Stellar Kinematics in AGNs}
\shortauthors{GREENE \& HO}

\begin{document}

\title{A Comparison of Stellar and Gaseous Kinematics in the Nuclei of 
Active Galaxies} 

\author{Jenny E. Greene}
\affil{Harvard-Smithsonian Center for Astrophysics, 60 Garden St., 
Cambridge, MA 02138}

\and

\author{Luis C. Ho}
\affil{The Observatories of the Carnegie Institution of Washington,
813 Santa Barbara St., Pasadena, CA 91101}

\begin{abstract}

To investigate the relationship between black holes and their host
galaxies, many groups have used the width of the \oiii\ $\lambda 5007$
line as a substitute for the stellar velocity dispersion (\sigmastar)
of galaxy bulges.  We directly test this assumption with a large and
homogeneous sample of narrow-line active galactic nuclei from the
Sloan Digital Sky Survey.  We consider multiple transitions (\oii\
$\lambda 3727$, \oiii\ $\lambda 5007$, and \sii\ $\lambda\lambda 6716,
6731$) and various techniques for quantifying the line width in order
to obtain a calibration between the gas velocity dispersion,
\sigmagas, and \sigmastar.  We find that \sigmagas\ of the
low-ionization lines traces \sigmastar, as does \sigmagas\ for the
core of \oiii\ after its asymmetric blue wing is properly removed,
although in all cases the correlation between \sigmagas\ and
\sigmastar\ has considerable scatter.  While the gas kinematics of the
narrow-line region of active galaxies are primarily governed by the
gravitational potential of the stars, the accretion rate, as traced by
the Eddington luminosity ratio, seems to play an important secondary
role.  Departures from virial motions correlate systematically with
accretion rate.  We discuss the implications of these results for
previous studies that use \oiii\ line widths to infer stellar velocity
dispersions in quasars and narrow-line Seyfert 1 galaxies.
\end{abstract}


\section{Introduction}

The recent discovery of a tight correlation between stellar bulge
velocity dispersion and central black hole (BH) mass (the \msigma\
relation) strongly indicates that BHs coevolve with their host
galaxies (Gebhardt \etal\ 2000a; Ferrarese \& Merritt 2000; Tremaine
\etal\ 2002).  The growth and accretion histories of BHs appear to be
inextricably linked with the mass assembly of galaxies.  Thus, the
evolution of BH mass density over cosmic time provides observational
constraints on galaxy evolution.  The \msigma\ relation allows us to
track BH mass density with time, provided we can measure a velocity
dispersion for a given system.

Unfortunately, the bulge stellar velocity disperion is often difficult
to measure directly.  In active galaxies the stellar absorption features are
quickly swamped by nonstellar emission from the nucleus.  There is
consequently a disconnect between the methods used to obtain BH masses
for active galactic nuclei (AGNs) and quiescent galaxies.  While
locally it is possible to infer a BH mass function 
based on the distribution of stellar velocity dispersion (e.g.,~Yu \&
Tremaine 2002), at significant distances we must rely on AGNs, which
are significantly brighter than stellar light.  The challenge
remains to develop secondary estimators of BH masses in AGNs and
assure ourselves that AGNs follow the same \msigma\ relation.

Luckily, there are a number of secondary techniques in development.
Reverberation mapping provides a virial estimate of BH mass to within
a factor that depends on the geometry of the broad-line region (BLR)
(e.g.,~Kaspi \etal\ 2000; Peterson \etal\ 2004).  This factor can be
empirically determined through comparison of \mbh\ inferred from
\sigmastar\ and the reverberation mapping mass (Gebhardt \etal\ 2000b;
Ferrarese \etal\ 2001; Nelson et al. 2004; Onken \etal\ 2004).
Reverberation mapping also provides an empirical calibration between AGN
luminosity and BLR radius, which may be combined with a BLR velocity
to yield a virial mass estimate (e.g.,~Kaspi \etal\ 2000).  This
technique has been used widely in the literature (e.g.,~McLure \&
Dunlop 2004) to study the cosmic evolution of BH mass density.

In addition to secondary mass estimates for the BH, there is a
secondary technique to estimate \sigmastar\ using the velocity
dispersion of the ionized gas (\sigmagas) from the narrow-line region
(NLR).  The NLR is in a privileged position in the galaxy: compact
enough to be illuminated by the active nucleus and yet large enough to
feel the gravitational forces of the bulge.  Early studies of
integrated line profiles determined that gravity, not nuclear
activity, dominates the global kinematics of the NLR (Wilson \&
Heckman 1985; Whittle~1992a,b). Most recently, Nelson \& Whittle
(1996) find \sigmagas\ $\simeq$ \sigmastar\ for a sample of 66 Seyfert
galaxies.  These studies all measure the width of \oiii\ $\lambda
5007$ because it is a strong and ubiquitous line.  It seems that
neither rotational support nor nuclear activity is important compared
to pressure support for the NLR, so that \sigmagas\ can be used as a
proxy for \sigmastar.  Of course, NLR gas velocity dispersion,
particularly of the \oiii\ line, is easier to measure for both active
and distant objects.  Many subsequent studies have used the full-width
at half maximum (FWHM) of \oiii\ as a proxy for \sigmastar.  Nelson
(2000) compares \mbh\ inferred from \sigmagas\ to reverberation-mapped
masses, while Boroson (2002) compares \sigmagas\ to virial masses
estimated using the line width-luminosity relation.  Shields \etal\
(2003) compare the cosmic evolution of BH mass to the evolution of
bulge velocity dispersion as inferred from \sigmagas, while both Grupe
\& Mathur (2004) and Wang \& Lu (2001) investigate the BH mass
distribution of narrow-line Seyfert 1 galaxies (NLS1s) using
\sigmagas.

Despite the widespread use of \sigmagas, caution is in order.  While,
to first order, the NLR kinematics seem to be determined by gravity,
the effects of the AGN are also apparent.  First of all, the \oiii\
line often has a broad blue wing (Heckman \etal\ 1981; De Robertis \&
Osterbrock 1984; Whittle 1985a; Wilson \& Heckman 1985 and references
therein) and Whittle (1992b) finds that powerful, compact linear radio
sources have anomalously broad \oiii\ lines.  At least in some cases
nuclear activity directly affects the NLR kinematics, and perhaps for
these systems a different emission line, one originating farther from
the central source, would be a better tracer of \sigmastar.  It is
also conceivable that rotation provides a significant source of
support for the gas in some fraction of NLRs, thus lowering \sigmagas\
relative to \sigmastar.  In short, it remains to be demonstrated with
a large and homogenous data set that \sigmagas\ truly traces
\sigmastar.  The data for such a study are now readily available from
the Sloan Digital Sky Survey (SDSS; York \etal\ 2000; Stoughton \etal\
2002).  Using the database presented by Brinchmann \etal\ (2004a) we
directly compare \sigmagas\ and \sigmastar\ for a large sample of
narrow-line AGNs from the SDSS.  In addition to the conventional
\oiii\ line, we consider the \oii\ $\lambda\lambda 3726, \ 3729$ and
\sii\ $\lambda\lambda 6716, \ 6731$ doublets. While these lines are
problematic due to blending, and are typically weaker than the \oiii\
line, they are less contaminated by strong asymmetries (as we show
below).  In addition, the \oii\ line is observable to higher redshift.

Ionized gas kinematics in the centers of galaxies is of broader
interest than simply as a BH mass indicator.  The structure and
kinematics of the gas relative to the stars may provide clues about
the dynamical history of the galaxy (e.g.,~Sofue \& Rubin 2001).  In many
cases, it seems that the gas rotates more slowly than the stars, in
both elliptical (e.g., Caldwell, Kirshner, \& Richstone 1986; Cinzano \&
van der Marel 1994) and spiral galaxies (Fillmore, Boroson, \&
Dressler 1986; Kent 1988; Kormendy \& Westpfahl 1989; Bertola \etal\
1995; Fisher 1997; Cinzano \etal\ 1999).  As noted by many of these
authors, this may indicate significant pressure support for the gas.
On the other hand, Vega Beltr\'{a}n \etal\ (2001) find
the gas in the centers of $\sim 20$ disk galaxies is moving on
circular orbits in a cold disk.  Presumably, a variety of dynamical
states are possible for the nuclear gas.  It would be interesting to
examine this question with a statistical sample such as SDSS.  In
particular, is there anything special about the dynamical state of the
ionized gas in AGNs?

We note that we are by no means the first to perform an analysis of
this type.  Earlier studies have looked in detail at both the \oiii\
line shape as an indicator of the kinematics of the NLR and at the
correlation of the line width with other nuclear and host properties
(for a review, see Wilson \& Heckman 1985 and references therein;
Nelson \& Whittle 1996).  The new aspect of this work is the large,
homogenous sample considered.  It is also, to our knowledge, the first
systematic investigation of the behavior of the \sii\ and \oii\ line
profiles.  Furthermore, this is the first systematic investigation of
NLR kinematics since the discovery of the \msigma\ relation.  Not only
does this provide new motivation for a better understanding of the
\sigmagas$-$\sigmastar\ relation, but it enables us to examine the
kinematics of the NLR in relation to the physical properties of the
AGN.  In particular, we find that the accretion rate of material onto
the BH partially determines the kinematics of the NLR.

\section{The Sample}

The Second Data Release (DR2) of the SDSS (Abazajian \etal\ 2004),
with 375,000 galaxy spectra, enables us to compare nuclear gas and
stellar kinematics with a large, homogeneous sample.  The reader is
referred to the project website for more details, as well as a
complete bibliography for the survey\footnote{http://www.sdss.org/}.
Briefly, spectroscopic candidates are selected based on imaging in the
$u$, $g$, $r$, $i$, and $z$ bands (Fukugita \etal\ 1996; Smith \etal\
2002; Strauss \etal\ 2002) with a drift-scan camera (Gunn \etal\
1998).  Spectra are acquired with a pair of fiber-fed spectrographs.
Each fiber subtends a diameter of $3\arcsec$, corresponding to $\sim
5.4$~kpc at $z=0.1$.  The spectra have an instrumental resolution of
$\lambda/\Delta \lambda \approx 1800$ ($\sigma_{\rm inst} \approx 71$
\kms).  Integration times are determined for a minimum signal-to-noise
ratio (S/N) of 4 at $g=20.2$ mag.  The spectroscopic pipeline performs
basic image calibrations, as well as spectral extraction, sky
subtraction, removal of atmospheric absorption bands, and wavelength
and spectrophotometric calibration (Stoughton \etal\ 2002).

Recently, the SDSS team at MPA/JHU made publicly available their database of 
galaxy properties for 211,894 galaxies, including 33,589 narrow-line (Type 2)
AGNs from DR2\footnote{http://www.mpa-garching.mpg.de/SDSS/}.  The
sample selection and spectral treatment are discussed in detail in
Kauffmann et~al. (2003a).  Like the main galaxy sample, the redshift
distribution is strongly peaked at $z \approx 0.1$, and H$\alpha$ is
always present in the band.  AGNs are selected based on their location in
line intensity ratio diagnostic diagrams (Baldwin, Phillips, \&
Terlevich 1981; Veilleux \& Osterbrock 1987; Kauffmann \etal\ 2003c)
after the stellar continuum is modeled and removed using stellar
population synthesis models.  Their definition of AGNs is very
inclusive.  Only objects solidly on the stellar locus are excluded,
while low-ionization nuclear emission-line regions (LINERs; Heckman 1980)
and transition objects (Ho, Filippenko, \& Sargent 1993) are included.
The database includes the line intensities used to make the selection.
They also make available stellar velocity dispersions, measured using
a direct-fitting algorithm that assumes Gaussian velocity profiles
(Heckman et~al. 2004).

We selected our sample from this set of narrow-line AGNs.  Using the
tabulated emission-line ratios, we use the line-ratio diagnostics
outlined in Ho, Filippenko, \& Sargent (1997a) to explicitly include
only Type 2 Seyferts, excluding both LINERs and transition objects.
This guarantees that the emission lines are dominated by AGN light.
As discussed in Ho (2004), the large physical scales subtended by the
SDSS fibers very likely includes substantial host galaxy emission,
which presents a source of confusion for interpreting LINERs and
transition objects.  We further require \oii, \oiii, and \sii\ to have a
S/N per pixel greater than at least 3 in the line, and a measured
stellar velocity dispersion.  The mean equivalent widths (EWs) for the
resulting sample are 18 for \oii, 20 for \oiii, and 4 for \sii, the weakest 
line.  In \S 3.4 we limit our sample to objects having only the highest-EW
lines to examine how the S/N affects our results.  
We exclude all objects with \sigmastar\ below the nominal SDSS instrumental
resolution of $\sigma_{\mathrm{inst}} \approx 70$~\kms\ (as in Heckman
\etal\ 2004).  We further exclude a handful of objects with
\sigmastar\ $> 400$~\kms\ as most likely unphysical (10 objects had
\sigmastar\ $\approx 600$~\kms)\footnote{The distribution function of
velocity dispersions for local early-type galaxies drops sharply for
\sigmastar\ $> 400$~\kms\ (Sheth et al. 2003).}  and 13 objects for
which one of the line detections appeared marginal after manual
inspection.  This left a sample of 2050 objects.  As described below,
we also remove objects with double-peaked or unresolved line profiles,
leaving a final sample of 1749 objects.  The median redshift of the
sample is $z \approx 0.1$, the same as the SDSS in general.

Throughout we assume the following cosmological parameters to derive
distances: $H_0 = 100h = 72$~\kms~Mpc$^{-1}$, $\Omega_{\rm m} = 0.3$,
and $\Omega_{\Lambda} = 0.7$.

\section{Data Analysis}


\begin{figure*}
\begin{center}
\epsfig{file=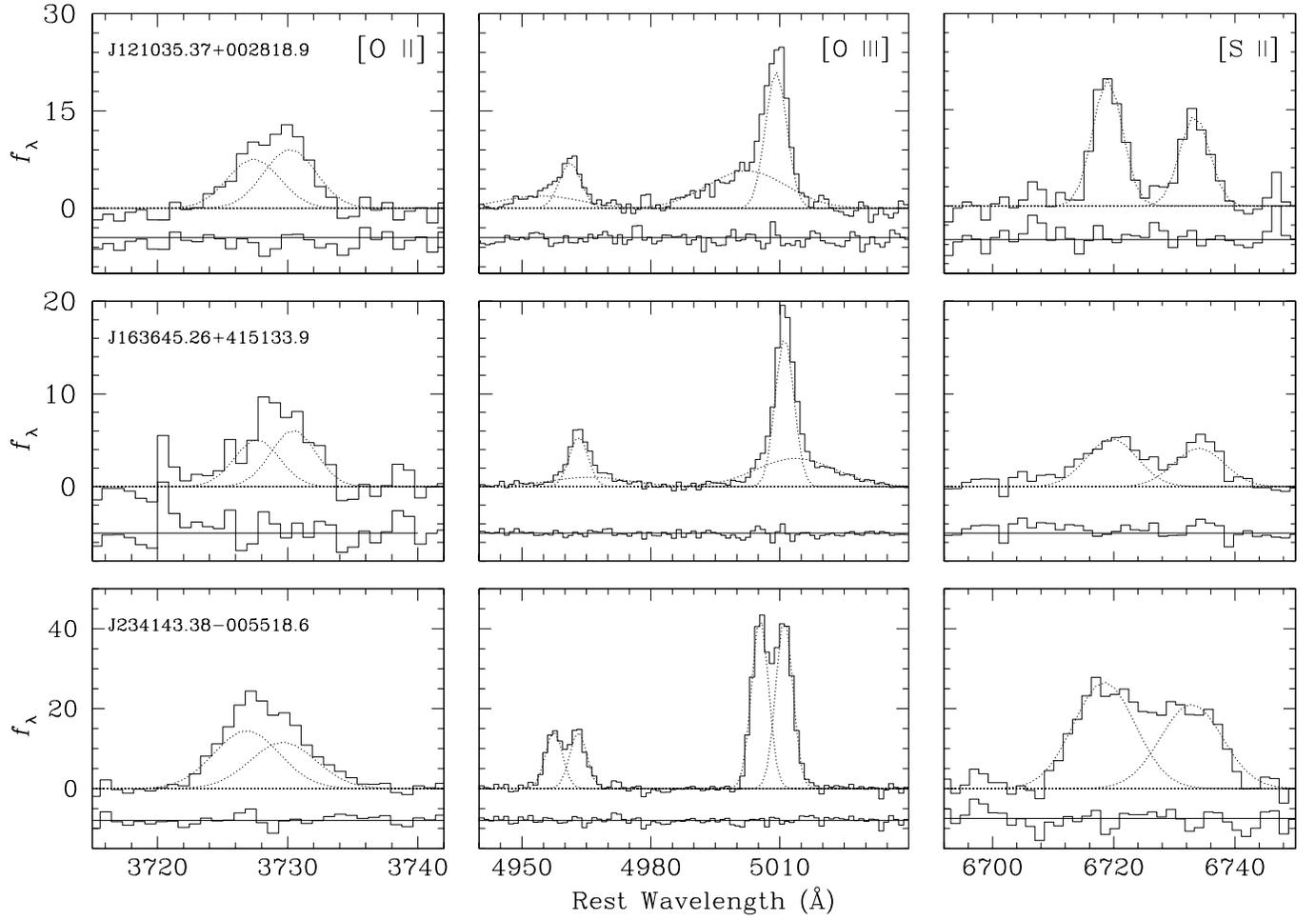,width=0.8\textwidth,keepaspectratio=true,angle=90}

\vskip -5mm
\figcaption[]{ 
Sample fits for blue-asymmetric ({\it
top}), red-asymmetric ({\it middle}), and double-peaked ({\it bottom})
objects.  The ordinate of the plots are in units of $10^{-17}$ erg
s$^{-1}$ cm$^{-2}$ \AA$^{-1}$.  The ordinate scales are 1:1:0.33 ({\it
top}), 1:3:1 ({\it middle}), and 1:1:0.2 ({\it bottom}).  Note the
hint of double-peakedness in the [S {\tiny II}] profile ({\it
bottom}).
\label{exspecfig}}
\vskip -5mm
\end{center}
\end{figure*}


\begin{figure*}
\begin{center}
\epsfig{file=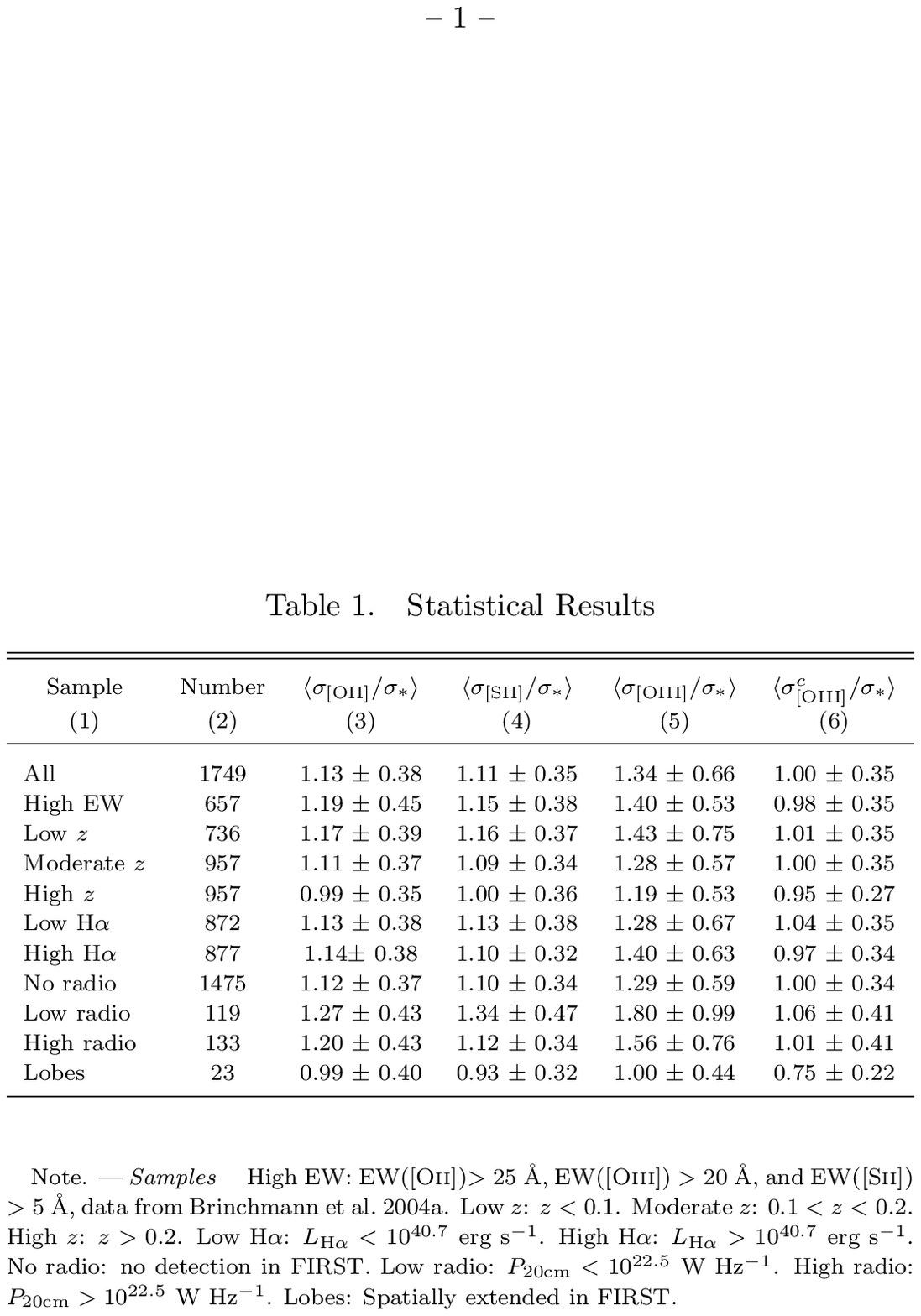,keepaspectratio=true}
\vskip -0.1mm
\end{center}
\end{figure*}


We wish to compare the stellar and gas line widths for the entire AGN
sample.  Since the stellar velocity dispersions have been measured, we
need only remove the stellar continuum and then measure the gas line
widths.  Below we describe each of those steps.

\subsection{Continuum Subtraction}

In order to measure reliable line widths, the underlying stellar
continuum must be removed.  Continuum variations can hide or mimic
broad line wings and skew relative line intensity measurements.
Since the MPA/JHU database does not provide continuum-subtracted spectra, we 
have to repeat this step of the analysis.
Fortunately, the SDSS survey provides a large and homogeneous library
of pure absorption-line galaxies for use in continuum subtraction.  We use a 
code that employs principal component analysis, kindly provided by Lei Hao 
(Hao \etal\ 2005).  A
library of hundreds of spectra is projected onto an orthogonal basis
of eigenspectra.  Typically eight eigen-components are sufficient to
provide an adequate model of the stellar spectrum.  A power-law
continuum and an A star spectrum may be included if necessary.

\subsection{Profile Fitting}

There are many possible ways to parameterize emission-line profiles.  Some of
the most common include the FWHM and the
second moment of the line, defined to be

$$\sigma^2 = ({\rm c}/\lambda_{\rm c})^2 
\int (\lambda - \lambda_{\rm c})^2 f_{\lambda} 
d\lambda \, / \int f_{\lambda} d \lambda .$$

\noindent
The line centroid is $\lambda_{\rm c}$,
$f_{\lambda}$ is the flux density in the continuum-subtracted
spectrum, and $\sigma$ is in units of \kms.
For a pure Gaussian
the second moment is equal to $\sigma$ and FWHM $= \sigma/2.35$.  
(The second moment is directly comparable to the stellar velocity dispersion
measurements, which assume a Gaussian broadening function.) If
to this original Gaussian we added a broad, low-contrast
Gaussian to represent the line wings, the FWHM would increase slightly,
by an amount depending on the contrast between the core and the wings. 
The second moment, on the other hand, is more sensitive to the presence of
wings, and becomes significantly broader.  For this reason, we get more
information on the full line profile from the second moment.
Unfortunately, measurements of the second moment can be very
sensitive to the S/N of the spectrum.  Therefore, we first fit a
multi-component Gaussian to the line profile, and then measure the
second moment from the model profile, as described below.  
This technique has the further benefit of providing a
decomposition of the core and wing components, so that 
we can examine the distribution of core and wing widths independently.

After much experimentation, we find that the most reliable method,
applicable to the majority of objects, is to fit each profile with two
Gaussians, one representing the line core and the other representing
the line wing. The second component
is fixed initially to be blueward of the central component, although
in a subset of cases ($\sim 6 \%$), a redshifted component was
required for an acceptable fit.  If the fit with a second component improves
significantly (at least $20 \%$ increase in \chisq)
over the single-component fit, it is kept.  We use this model to
simultaneously fit the \oiii~$\lambda \lambda$4959, 5007 lines, with
their separation fixed and their relative strength constrained to
the theoretical ratio of 3:1.  Likewise, the \sii~$\lambda
\lambda$6716, 6731 lines are fixed to the same profile\footnote{If the 
NLR is stratified in density, the width of each component of the doublet can 
be different, since their critical densities are different.  
But the profile differences are expected to be small and hard to discern 
with data of moderate S/N.}, with their separation
fixed to the laboratory value;  the ratio of the
line intensities varies with density, and is allowed to vary in the fit.
The \oii~$\lambda \lambda$3726, 3729 lines behave similarly, but the
decomposition is more difficult because the expected line widths are
comparable to the pair separation (2.8 \AA\ or 225 \kms).  For this
reason, we allow the line ratio to vary between 0.2 and 1.4, the expected
range for the typical densities observed in the NLR
(e.g.,~Osterbrock 1989). Sample fits are shown in Figure 1.

As in previous studies, the majority of the line profiles are smooth,
but a small percentage are lumpy, and some have two distinct peaks
with comparable flux and line width.  We find 31 of the 2050 ($\sim 1
\%$) objects with clearly defined double peaks (Fig. 1).  While we
show below that in general the NLR is not rotationally dominated, we
suspect that in these cases the NLR gas is predominantly organized
into a disk.  Of course, bipolar outflows may also be responsible for
the double-peaked lines.  In any case, these objects are such clear outliers
that we exclude them from further consideration.

\subsection{Instrumental Resolution}

The observed $\sigma$ is the convolution of the true line width and
the instrumental response.  To first order, the intrinsic line width
can be approximated by $\sigma = \sqrt{\sigma_{\rm obs}^2 -
\sigma_{\rm inst}^2}$.  In the case of the SDSS data, $ \sigma_{\rm
inst}$ varies not only as a function of wavelength, but also the
location of the object on the plate and the temperature on the night
of the observations.  For this reason, the SDSS pipeline measures the
instrumental response using an arc lamp spectrum, and returns the
resolution at every pixel.  The mean values for this sample are
$74\pm8$ \kms\ for \oii, $60\pm5$ \kms\ for \oiii, and $56\pm5$ \kms\
for \sii.  When we remove the effects of instrumental resolution, some
of the objects become unresolved.  Measurements below the resolution
limit are not reliable, and so we remove all objects for which one or
more of the lines in question is unresolved, leaving a total of 1749
objects.

\subsection{Errors and Uncertainties}

The average formal error in \sigmagas\ is $\sim 5 \%$.
However, this estimate neglects systematic effects, the two most
serious of which are uncertainties in the continuum subtraction and 
component decomposition.  The latter is only a concern in the
case of \oiii, the only line that often requires a two-component
fit. In some cases, there is ambiguity as to the
exact decomposition between these two.  However, we do not believe
errors in decomposition introduce any systematic bias.
There are publicly available error 
estimates\footnote{http://spectro.princeton.edu/} for the stellar velocity
dispersions we are using.  The median error is $\sim 8 \%$, or $\sim
12$ \kms\ at the median \sigmastar\ $\approx 144$ \kms.  These
uncertainties are small compared to the scatter in \meansig, the quantity 
central to our analysis in this paper.

We can investigate the extent to which the scatter is due to
measurement error by limiting the sample to a subset of high-EW
objects for which the formal measurement errors are small.  We limit
the sample to the strongest $\sim 500$ objects for each line.  We find
no significant difference in either the mean value or the scatter
around \sigmagas\ (see entries in Table 1).  This suggests that the
scatter is dominated by intrinsic variations, rather than measurement errors.
In the Appendix we specifically examine the scatter
introduced by rotational broadening.  We find that it is also
insufficient to account for the observed scatter.

\section{Results}

\subsection{Primary Driver: Gravitational Potential of the Bulge}

Our findings are summarized in Table 1, which shows \meansig\ for each
transition.  Our primary result is that the line widths of \sii\ and
\oii\ track the stellar velocity dispersion in the mean, albeit with
substantial scatter. The width of \oiii, on the other hand, is on
average significantly broader than \sigmastar; $\sigma_{\mathrm{[O
{\tiny III}]}}$ cannot be used directly as a tracer of \sigmastar.
However, as we discuss below, the core of \oiii\ does track
\sigmastar.

The different behavior of the low-ionization lines compared to \oiii\
can be understood in terms of well-known variations in line shape as a
function of critical density and ionization potential within the NLR
(e.g.,~Pelat, Alloin, \& Fosbury 1981; Filippenko \& Halpern 1984;
Filippenko 1985; De Robertis \& Osterbrock 1986).  The NLR is
stratified in density such that lines of higher critical density
originate from closer to the nucleus. Since the critical density of
\oiii\ $\lambda 5007$ is higher than those of \oii\ $\lambda 3727$ and
\sii\ $\lambda\lambda 6716, 6731$ (Osterbrock 1989), in a
density-stratified NLR we expect \oiii\ to be broader than \oii\ and
\sii, whereas \oii\ and \sii\ should share similar line widths.  As
expected, we find that $\langle \sigma_{\mathrm{[O {\tiny
III}]}}/\sigma_{\mathrm{[O {\tiny II}]}} \rangle$ and $\langle
\sigma_{\mathrm{[O {\tiny III}]}}/ \sigma_{\mathrm{[S {\tiny II}]}}
\rangle \approx 1.2 \pm 0.5$, while $\langle \sigma_{\mathrm{[S {\tiny
II}]}}/\sigma_{\mathrm{[O {\tiny II}]}} \rangle \approx 1.0 \pm 0.3$.
This suggests that the \oiii\ emission originates from closer to the
nucleus than the other lines.

As noted, our results suggest that the kinematics of low-ionization
gas in the NLR is dominated by gravity, such that the velocity
dispersion in the gas and stars are comparable.  On the other hand, the
\oiii\ line width is significantly broader than \sigmastar, suggesting
that the active nucleus plays a significant role in governing the
kinematics of the \oiii-emitting gas.  Taken at face value, this
finding seems to contradict the conclusions of Nelson \& Whittle
(1996), who have argued that the \oiii\ line width is largely
dominated by gravity.  Since the \oiii\ profile is invariably
asymmetric, characterized by a central core component plus a
high-velocity blue wing, it is reasonable to suppose that the line
core should have the closest affinity to any gravitationally dominated
component.  Using our two-component Gaussian fits to the \oiii\
profiles, we can directly compare the core velocity dispersion with
\sigmastar.  The result of this test is shown in the last column of
Table 1.  Indeed, we find that when the blue wing is removed, the core
of the \oiii\ line statistically traces \sigmastar, again with
significant scatter.  This suggests that the low-velocity gas in the
line core is gravitationally dominated, while the wing is more
strongly influenced by the active nucleus.  

\subsection{Secondary Drivers}


\begin{figure*}
\begin{center}
\epsfig{file=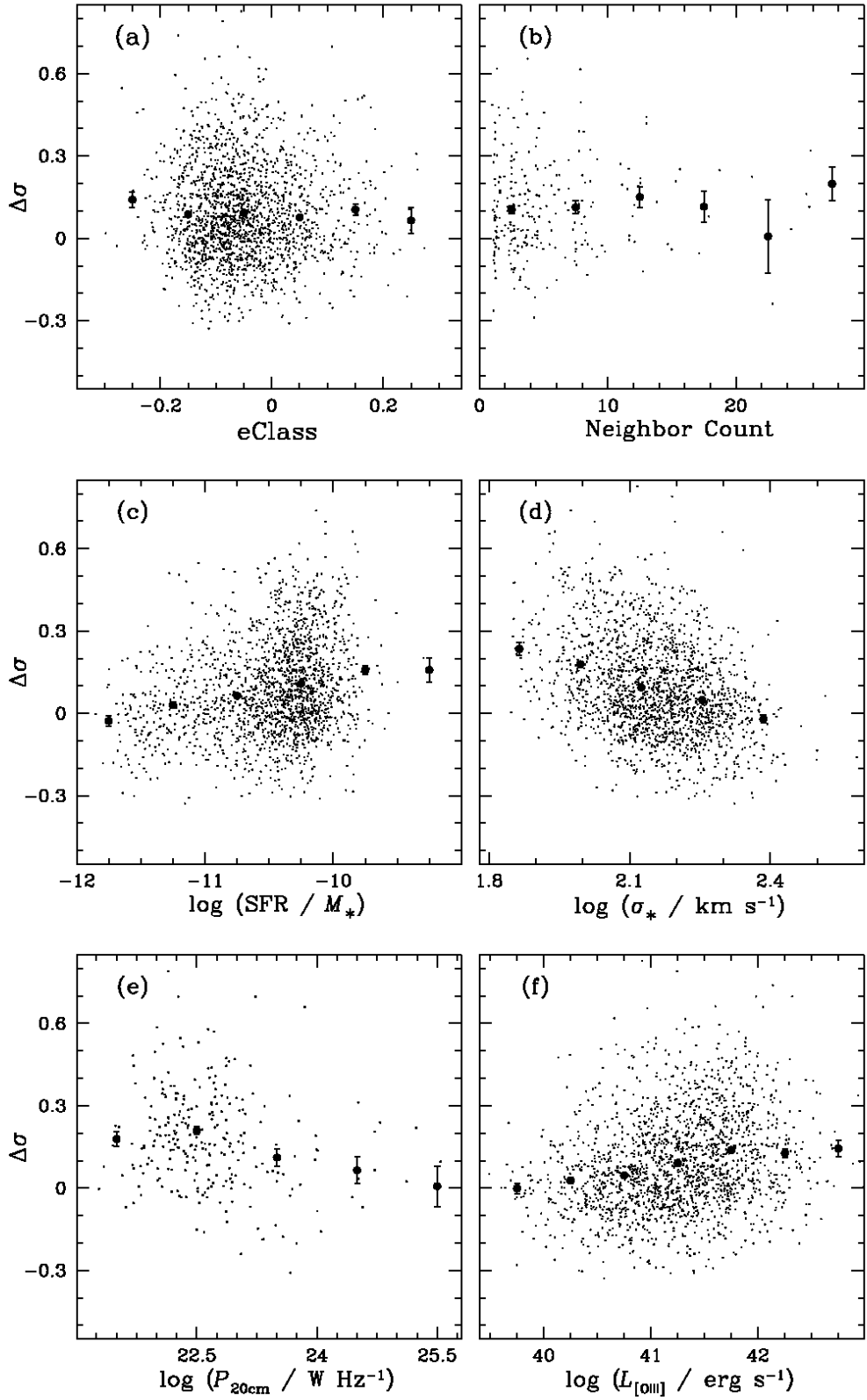,width=0.95\textwidth,totalheight=8.5in,keepaspectratio=true}
\vskip -0.1mm
\figcaption[]{ 
Deviations from virial line width, 
\delsig\ $\equiv \log \sigma_{\mathrm{[O {\tiny III}]}} - \log
\sigmastar$, 
with various parameters:
({\it a}) eClass,
({\it b}) neighbor count,
({\it c}) median total star formation rate normalized to
the galaxy stellar mass (yr$^{-1}$),
({\it d}) \sigmastar,
({\it e}) radio power, and
({\it f}) [O {\tiny III}] luminosity.
In all cases, filled circles show the mean value plotted in the 
middle of the bin, and error bars are the standard deviation in the mean.
\label{delfig}}
\vskip -5mm
\end{center}
\end{figure*}


We now look for secondary parameters to explain the deviations of
\sigmagas\ for the \oiii\ line compared to \sigmastar, which are
plausibly related to either the active nucleus or the environment.
Looking at the correlation of these deviations with various indicators
of nuclear activity may reveal useful information about the
interaction of the nucleus with the NLR.  We consider host galaxy
morphology, local environment, star formation rate, and bulge stellar
velocity dispersion as global variables.  In terms of nuclear effects,
we look at radio power, AGN luminosity, and Eddington ratio [\lledd,
where $L_{\mathrm{Edd}} \equiv 1.26 \times
10^{38}$~erg~s$^{-1}$~(\mbh/\msun)].  In
particular, we will examine the correlation of each of the above
properties with \delsig\ $\equiv \log \sigma_{\mathrm{[O {\tiny
III}]}} - \log \sigmastar$.  In Figure 2 we show \delsig\ plotted
against each variable, with binned values overplotted to highlight any
underlying trend.  The large, binned points represent the mean and
standard deviation in the mean plotted at the bin center.
 
Of additional interest is whether the characteristics of the blue wing
correlate with any of the properties listed above.  In order to
investigate this question we looked at three wing diagnostics: the width of
the blue wing, the flux ratio between the core and the wing, and the
wavelength shift between the core and the wing.  We found no strong
correlations between any of the above quantities with either the
nuclear or global variables.  

\subsubsection{Global Properties}

Both gas and stellar kinematics in the central regions of galaxies may
depend on Hubble type.  We cannot derive robust morphologies directly
from the SDSS data because the resolution and depth are insufficient.
However, Shimasaku \etal\ (2001) and Strateva \etal\ (2001) find that
the color, concentration index, and spectral decomposition provided by
the SDSS pipeline all correlate well with standard morphological
classifications.  We use two different integrated properties to
estimate the morphological distribution of the sample.  The ``eClass''
is a spectroscopic classification derived from the principal component
analysis decomposition of each galaxy spectrum, and ranges from
$-0.35$ for early-type galaxies to +0.5 for late-type galaxies.  The
``concentration index,'' the ratio of the radii enclosing $90 \%$ and
$50 \%$ of the light in the $r$ band, describes the compactness of the
stellar emission.  Concentrations indices larger than $2.6$ correspond
to early-type galaxies.  The mean concentration index of 2.65 and mean
eClass of $-0.05$ are both consistent with a population of early-type
galaxies (Kauffmann \etal\ 2003c).  We see no trend with \delsig\ in
either of these parameters (see Fig. 2{\it a}\ for \delsig\
vs. eClass).

Nelson \& Whittle (1996) find evidence for broadened line widths in
noticeably disturbed systems.  While we cannot use the SDSS imaging to
reliably gauge disturbance levels, we can
look for systematic trends with local galaxy density.
Kauffmann \etal\ (2004) find an increase of nuclear activity and
star formation for AGNs in overdense regions.  We might, therefore,
expect some correlation between environmental density and line width.
However, we find no evidence for a trend.  
The luminosity-weighted neighbor counts are
plotted against \delsig\ for those objects with measured neighbor
counts in Figure 2{\it b}.  

Depending on the origin of the ionized gas, star formation
(particularly in the nuclear regions) could affect the gas kinematics.
Recent studies, such as that of Kauffmann \etal\ (2004), suggest a
link between star formation and nuclear activity.  If this is the
case, then stellar winds from massive stars or supernova feedback may
impact the dynamical state of the nuclear gas. To investigate this
possibility, we use the total star formation rate per unit stellar
mass (SFR/$M_*$) from Brinchmann \etal\ (2004b), which is derived for
the AGN sample using a relation between the 4000 \AA-break and SFR in
a sample of star-forming galaxies.  In Figure 2{\it c}\ we plot log
SFR/$M_*$ against \delsig; there is no significant trend apparent.

By far the strongest trend is between \delsig\ and \sigmastar\
(Fig. 2{\it d}), with a Spearman rank correlation coefficient of $\rho
= -0.32$, and a probability of $P_{\rm null} < 10^{-4}$ for
the null hypothesis of no correlation.  Massive galaxies with large
values of \sigmastar\ on average have \delsig\ $\approx$ 0, whereas
less massive galaxies with smaller values of \sigmastar\
systematically have \delsig\ $> $ 0.  This trend with \sigmastar\
directly translates into a trend with BH mass, via the \msigma\
relation.

\subsubsection{Radio Power}

Many lines of evidence suggest that radio jets influence the physical
and dynamical state of the NLR.  Wilson \& Willis (1980) found a
correlation between \oiii\ line width and radio power.  There is also
a well-known relation between \oiii\ luminosity and radio power (de
Bruyn \& Wilson 1978; Ho \& Peng 2001).  Whittle (1985b,1992a,b) finds
evidence that strong ($L_{1415~\mathrm{MHz}} \gtrsim 10^{22.5}$ W
Hz$^{-1}$), ``linear''\footnote{The ``linear'' classification for the
above studies comes from the high-resolution (0\farcs2--0\farcs6 or
80--400 pc) study of Ulvestad \& Wilson (1984).  Linear sources have
extended radio emission on sub-galactic scales (tens to hundreds of
parsecs), as opposed to sources classified as ``diffuse'', ``slightly
resolved'', or ``unresolved.''  The linear morphology suggests the
presence of compact jets, not to be confused with the extended jets
and radio lobes found in radio galaxies.} radio sources have
systematically broader \oiii\ lines.  The same effect is seen by
Nelson \& Whittle (1996), but the trend is weaker.  It seems that the
jet plasma interacts with and accelerates the NLR, thus boosting the
line width.  We stress that this finding holds only for compact linear
radio sources.  In contrast, Smith, Heckman, \& Illingworth (1990) do
not find supervirial \oiii\ line widths among powerful radio galaxies
with kpc-scale jets.

In light of the above results, we extracted radio fluxes from the Faint
Images of the Radio Sky at Twenty-cm (FIRST; Becker, White, \& Helfand
1995) survey, which is largely coincident with the SDSS footprint.
FIRST is a 20 cm survey with a sensitivity of $\sim$ 1 mJy and a
resolution $\sim 5$\arcsec.  All but 279 of the sources in our sample have 
FIRST measurements.  The SDSS pipeline only considers matches within
2\arcsec\ in order to minimize confusion (Ivezi\'{c} et~al. 2002).
However, a small fraction of the sources will have extended sources
associated with them as well.  We therefore search the FIRST database
with a 60\arcsec\ radius for extended sources, and find 25.  The radio
luminosities are K-corrected to 20 cm assuming $F_{\nu} \propto
\nu^{-0.46}$, which is the median spectral slope found by Ho \&
Ulvestad (2001) for nearby Seyfert galaxies.  Unfortunately, while the
FIRST database is homogeneous and complete, the resolution is not
sufficient to resolve nuclear morphology.  Nevertheless, for
completeness we consider possible trends with \delsig.

Our results are ambiguous (see Fig. 2{\it e}).  As can be seen from
Table 1, the presence of core radio emission seems to have no impact
on the observed line widths.  On the other hand, extended radio
sources appear to have marginally {\it narrower} line widths.  We
attribute the ambiguous results to the low sensitivity and resolution in
the radio data.  The correlation previously discovered holds for
linear radio cores only.  With 5\arcsec\ resolution, we cannot
discriminate between different core morphologies.  Furthermore, the
sensitivity limit of $\sim 1$ mJy corresponds to $P_{20\mathrm{cm}} =
2.5\times 10^{22}$ W Hz$^{-1}$, at the ``powerful'' limit considered
by Whittle (1992b).

\subsubsection{AGN Luminosity and Eddington Ratio}

Perhaps the most basic question to ask is whether the energy output of
the active nucleus is directly related to \delsig.  We use
emission-line (\halpha\ and \oiii) luminosity as a surrogate for the
AGN luminosity (Adams \& Weedman 1975; Yee \& Oke 1978; Yee 1980).  We
derive \lha\ and \loiii\ from the fluxes given in the database of
Brinchmann \etal\ (2004a), corrected for Galactic
extinction using the dust maps of Schlegel, Finkbeiner, \& Davis
(1998).  We find no evidence for a strong trend with
either quantity (see Fig. 2{\it f}\ for \loiii),
at least for the relatively limited range of luminosities probed here.
This result does not change for a high-EW subsample.  


\begin{figure*}
\begin{center}
\vskip -40mm
\epsfig{file=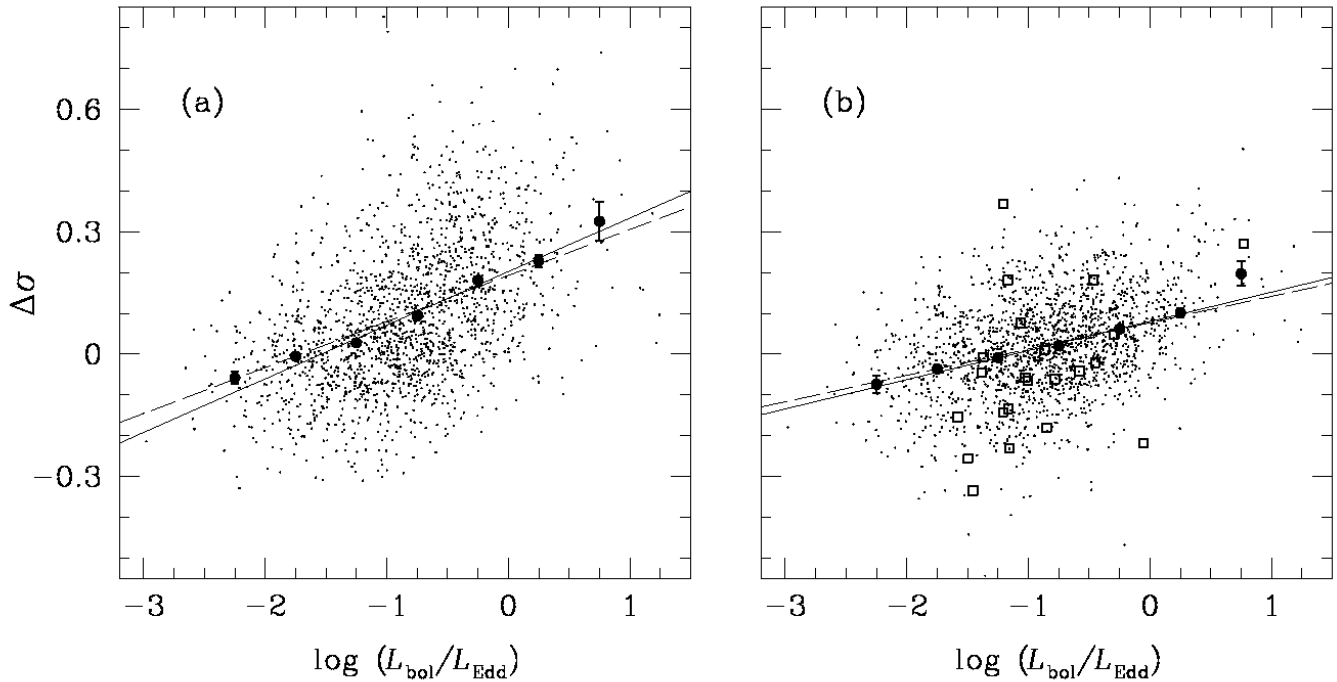,width=0.78\textwidth,keepaspectratio=true,angle=-90}
\vskip -15mm
\figcaption[]{ 
({\it a}) 
\delsig\ $\equiv \log \sigma_{\mathrm{[O {\tiny III}]}} - \log
\sigmastar$ plotted against the Eddington ratio, $L_{\rm
bol}/L_{\rm Edd}$.  Bolometric luminosity is estimated using the
bolometric correction from Heckman \etal\ (2004); see text. Points are
the SDSS sample.  Large points are the mean values plotted in the
center of the bin, and error bars are the standard deviation in the
mean.  Solid line shows the fit to data in Equation 1; dashed line
shows the fit to the binned data in Equation 2.  ({\it b}) Same as
({\it a}), but using \sigmagas\ = FWHM/2.35.  Boxes are the Type 1
objects from the Nelson \& Whittle (1996) sample.
The solid and dashed lines represent Equations 3 and 4, respectively.
\label{ledfig}}
\vskip -8mm
\end{center}
\end{figure*}


In addition to the luminosity of the AGN, we can ask whether the Eddington
ratio affects the NLR kinematics.  The Eddington
ratio, proportional to the accretion rate per unit (BH) mass, characterizes 
the extent to which radiation pressure competes with gravity in the
nucleus.  Therefore, we might expect objects with higher Eddington ratios to
show larger nonvirial motions.

We do not have direct measurements of Eddington ratios for our sample,
but we can estimate them by using line luminosities as a substitute
for \lbol\ and \sigmastar\ as a surrogate for BH mass (and hence
\ledd).  We stress, however, that the estimated values of \lledd\ are
highly uncertain.  The conversion from the observed luminosity in a
specific band (e.g., optical) to bolometric luminosity introduces the
largest uncertainty.  For Type 2 objects, the problem is compounded by
the fact that the nonstellar (AGN) continuum is not directly
observable in the optical.  Instead, we will make use of the
extinction-corrected \oiii\ luminosity to estimate $L_{\rm bol}$,
using the bolometric correction for \oiii\ luminosity from Heckman
\etal\ (2004).  This is derived in two steps.  First, the \oiii\
luminosity is correlated with optical AGN continuum luminosity for a
sample of Type 1 objects in SDSS (Kauffmann \etal\ 2003c; Zakamska
\etal\ 2003).  Then a standard bolometric correction from optical
continuum to bolometric luminosity (from Marconi \etal\ 2004) is
employed.  The final derived relation becomes $L_{\mathrm{bol}}
\approx 3500 L_{\mathrm{[O \tiny III]}}$, with an uncertainty of 0.4
dex.  We note that this estimate agrees well with the \oiii\
bolometric correction derived by L. C. Ho (2005, in preparation),
established in a completely independent way using broad-band spectral
energy distributions.  The BH mass is calculated using the measured
\sigmastar\ and the \msigma\ relation of Tremaine \etal\ (2002), which
formally has a scatter of less than 0.21 dex.

Notwithstanding the large uncertainty in bolometric correction, we
find a highly significant correlation between $\Delta \sigma$ and \lledd\ 
(Fig. 3{\it a}).  {\it The \oiii\ line width is
most supervirial in objects at the highest Eddington ratio.}
Formally, the Spearman rank correlation coefficient is $\rho = 0.46$,
with a probability of $P_{\rm null} < 10^{-4}$ for rejecting the null
hypothesis of no correlation.  For the purpose of comparing with
previous work, we also calculate \delsig\ using FWHM/2.35 rather
than \sigmagas.  Although we have argued that the FWHM is an
inferior measure of line width, 
there are occasions in which only FWHM measurements
are available in the literature.  The result is shown in Figure 3{\it
b}.  While the slope is shallower, the trend persists ($\rho = 0.36$,
$P_{\rm null} < 10^{-4}$).  

We have seen that \delsig\ is also significantly correlated with 
\sigmastar, or, equivalently, BH
mass (\S 4.2.1).  Since \lledd\ depends on BH mass, we must
consider whether BH mass or \lledd\ is really the
fundamental parameter driving \delsig.  Because the correlation of
\delsig\ with \lledd\ ($\rho = 0.46$) is stronger than that with BH mass
($\rho = -0.32$), \lledd\ appears to influence \delsig\ more strongly
than BH mass.  In principle, the strongest correlation could come from
some combination of the BH mass and \lledd.  However, we find that
including the BH mass in a partial correlation analysis (Akritas \&
Siebert 1996) does not improve the correlation.  We therefore conclude
that \lledd\ is the primary physical parameter driving \delsig.

The solid line in Figure 3{\it a}\ represents our fit to the data.
Using an ordinary least-squares regression with log \lledd\ as the
independent variable, we find 

\begin{equation}
\Delta \sigma = (0.131 \pm 0.006)~\mathrm{log}~
L_{\mathrm{bol}}/L{\mathrm{_{Edd}}} + (0.202 \pm 0.006).  
\end{equation}

\noindent
To better highlight the trend, we also binned the data in bins of 
$\Delta$(log \lledd) = 0.5 dex, which yields the following fit (dashed line):

\begin{equation}
\Delta \sigma = (0.113 \pm 0.005)~\mathrm{log}~
L_{\mathrm{bol}}/L{\mathrm{_{Edd}}} + (0.192 \pm 0.008).  
\end{equation}

\noindent
If we calculate \delsig\ using \sigmagas\ = FWHM/2.35 (Fig. 3{\it b}), then 
the corresponding relations for the unbinned and binned data are, 
respectively,

\begin{equation}
\Delta \sigma = (0.072 \pm 0.005)~\mathrm{log}~
L_{\mathrm{bol}}/L{\mathrm{_{Edd}}} + (0.080 \pm 0.005)
\end{equation}

\noindent
and

\begin{equation}
\Delta \sigma = (0.065 \pm 0.002)~\mathrm{log}~
L_{\mathrm{bol}}/L{\mathrm{_{Edd}}} + (0.077 \pm 0.004).  
\end{equation}

It is interesting to examine \delsig\ at the extreme Eddington ratios.
For objects at the lowest Eddington ratios (log \lledd\ \lax $-2$ for
our sample) we find line widths that are {\it subvirial}\ ($\Delta
\sigma < 0$).  This finding suggests that quiescent (inactive)
early-type galaxies (most closely approximated by the lowest-Eddington
ratio systems) have subvirial gas velocities.  In support of this
finding, Vega Beltr\'{a}n \etal\ (2001) find the \sigmagas\ $<$
\sigmastar\ in early-type spirals, where the gas is at least partially
rotationally supported (in contrast with late-type spirals, where both
gas and stars are rotationally supported).  In light of this issue, it
is interesting to note that the lobe-dominated radio sources, which
have been suggested to be preferentially low-Eddington ratio objects
(e.g.,~Rees \etal\ 1982), also appear to have subvirial gas velocities
(\S~4.2.2).  We do not find a significantly low distribution of
Eddington ratios among the extended radio sources in our sample, but
the statistics are poor.  It would be interesting to extend our study
to a larger sample of extended radio sources.

At the other extreme, for log \lledd\ \gax\ 1, our results imply that
\sigmagas\ systematically overestimates \sigmastar\ by $\sim 0.2-0.3$
dex.  This strongly suggests that studies that use \sigmagas\ as a
proxy for \sigmastar, such as that by Shields \etal\ (2003) or Grupe
\& Mathur (2004), may systematically overestimate \sigmastar.  In both
cases we expect the objects to be preferentially accreting at a high
rate.  Shields et al. (2003) concentrated on high-luminosity sources
(QSOs) at high redshift, which would tend to be biased toward higher
Eddington ratios.  The study of Grupe \& Mathur (2004) is concerned
with NLS1s, which are thought to be preferentially high-Eddington ratio
objects (e.g.,~Pounds, Done, \& Osborne 1995).  Grupe \& Mathur (2004;
see also Mathur, Kuraszkiewicz, \& Czerny 2001), using
FWHM$_{\mathrm{[O \tiny III]}}$ as a surrogate for \sigmastar, claim
that NLS1s have BHs that are undermassive relative to their bulge
masses.  Most or all of the trend discussed by these authors may
be attributed to the systematic broadening of \oiii\ at the highest
Eddington ratios.

In Figure 4 we show a modified version of Figure 1 from Grupe \&
Mathur (2004).  We have reproduced the figure exactly in open symbols,
and then overplotted the same data with the \oiii\ widths corrected
using our derived calibration (Eq. 3) in filled symbols. Again, while
we do not advocate the use of FWHM, in this case it is the only
available measure of line width.  The squares are ``normal''
broad-line Seyfert 1s, while the circles are NLS1s, selected using the
Grupe \& Mathur (2004) criterion of FWHM(\hbeta) $\leq$ 2000 \kms.
The \mbh\ values are derived from FWHM(\hbeta) and \lf\ as tabulated
in Grupe \etal\ (2004; corrected to our cosmology), using the line
width-luminosity-mass scaling of Kaspi \etal\ (2000).  The overplotted
line represents the Tremaine \etal\ (2002) parameterization of the
\msigma\ relation.  The \lledd\ values used for the correction are
derived using the $L_{\mathrm{bol}}$ values provided in Grupe \etal\
(2004), estimated from their optical and soft X-ray data.  The mean
Eddington ratio for the broad-line Seyfert 1s in this sample is log
\lledd\ = $-0.74$, while for the NLS1s it is log \lledd\ = 0.23, in
support of the idea that NLS1s tend to be high-Eddington ratio
objects.  This translates into $\langle$\delsig $\rangle = 0.03$ for
the broad-line objects, as compared to $\langle$\delsig $\rangle =
0.1$ for the NLS1s.  As shown in Figure 4, the argument that the NLS1s
lie systematically below the \msigma\ relation is significantly
weakened.  To test this statistically, we compute the perpendicular
distance ($d_\perp$)\footnote{If $d_x$ is the difference between the
observed [O{\tiny III}] line width and the Tremaine et al. (2002)
prediction based on the \mbh\ measurement, and $d_y$ is the difference
between the observed \mbh\ and that predicted by the measured [O{\tiny
III}] line width, then $d_\perp$ = $d_x$ $d_y$ / $\sqrt{d_x^{2} +
d_y^{2}}$.  We (arbitrarily) define $d_\perp > 0$ if the point lies
below the Tremaine \etal\ (2002) line.  We choose this statistic
because it does not assume an independent variable.} of each point
from the Tremaine \etal\ (2002) line.  After our correction, the two
populations are no longer statistically different in $d_\perp$
according to a Kolmogorov-Smirnov (K-S) test ($P_{\rm KS}$ = 25\%).
In other words, the case that the BHs of NLS1s are undermassive with
respect to their bulges is no longer statistically supported.

We are wary to overinterpret Figure 4, since it is based on our
statistical correction for \delsig\ and on a bolometric correction,
both of which inherently have a large scatter.  In actuality, the best
way to establish the location of NLS1s relative to the \msigma\ relation
is to obtain direct measurements of \sigmastar\ for them.  Barth,
Greene, \& Ho (2005), in a systematic investigation of \sigmastar\ for
the NLS1s selected from the sample of Greene \& Ho (2004), show that
these objects follow the \msigma\ relation.

\section{Discussion}

\subsection{Seyfert Type}

We must examine the extent to which the results presented here are
applicable to broad-line (Type 1) AGNs, since it is precisely in these
objects that narrow emission-line widths are used to estimate
\sigmastar.  According to the unification hypothesis of Seyfert
galaxies, the differences between Type 1 and Type 2 sources arise from
orientation alone (e.g.,~Antonucci \& Miller 1985), such that the BLR
in Type 2 AGNs is hidden from direct view by an obscuring torus.  In
this picture, the NLR is an extended, roughly isotropic region
extending far beyond the torus.  Thus, our results should apply to
both AGN types.  On the other hand, it is possible that the fractional
incidence of blue wings depends on the obscuring geometry, which in
turn would depend on Seyfert type.  This could result, for instance,
if the torus is partially responsible for collimating the ionizing
radiation field (see, e.g.,~Tadhunter \& Tsvetanov 1989).  That said,
to first order at least, our results should apply to Type 1 and Type 2
objects.  We offer two lines of evidence to support this.

First, if we consider well-studied, unbiased samples of Seyfert
galaxies, such as that derived from the Palomar survey

\epsfig{file=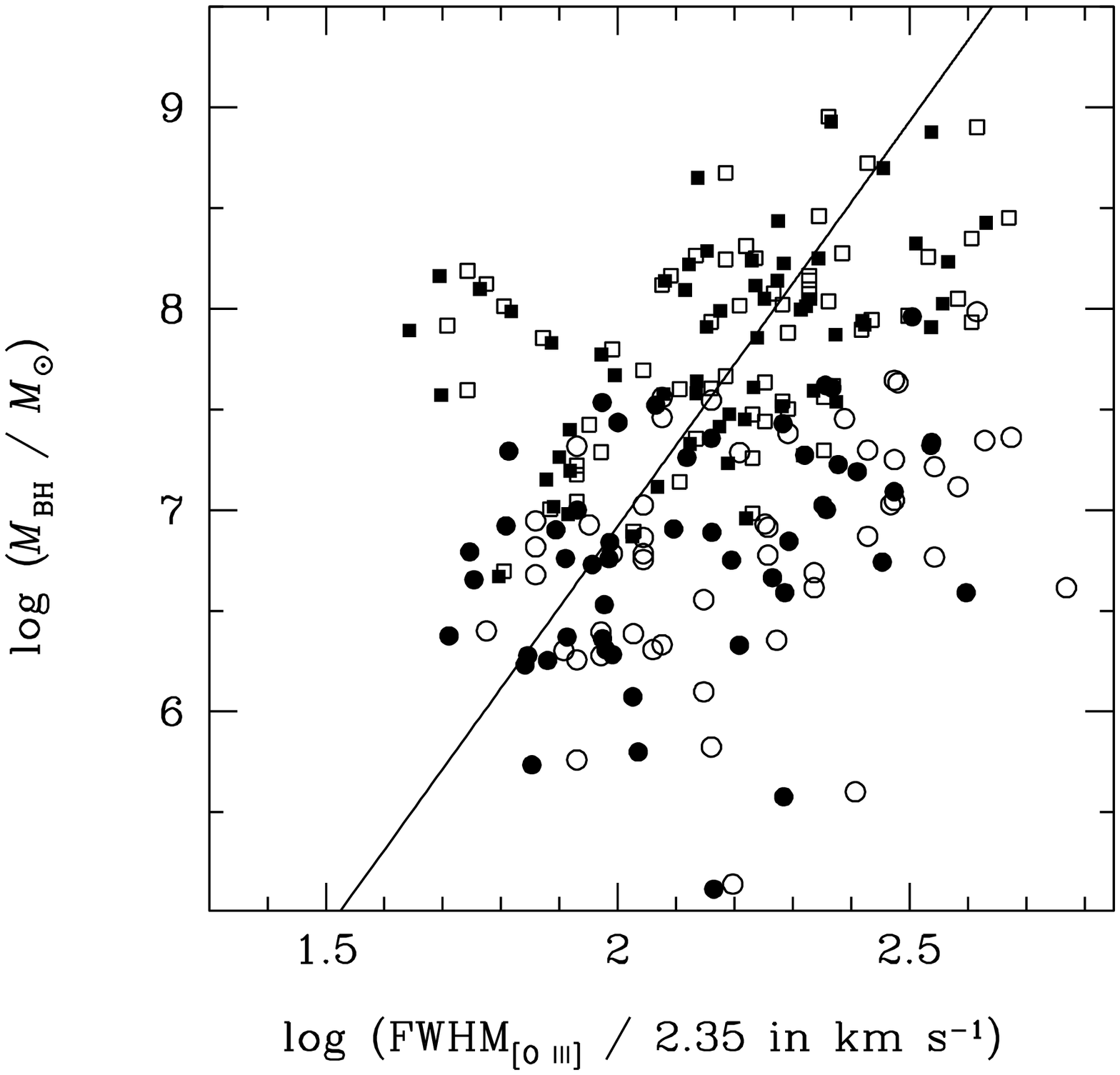,width=0.5\textwidth,keepaspectratio=true,angle=0}
\vskip -0.1mm
\figcaption[]{ 
\mbh\ vs. [O {\tiny III}] line width for the Grupe \& Mathur (2004)
sample.  We use the FWHM in this case because it is the quantity published.
Open symbols are a reproduction of Figure 1 in Grupe \&
Mathur (2004), while filled symbols show the data for
$\sigma_{\mathrm{[O {\tiny III}]}}$ corrected for the Eddington ratio
according to Equation 3.  Circles are NLS1s; boxes are broad-line
Seyfert 1s.  All measured quantities were taken from Grupe \etal\
(2004).  The \mbh\ values come from the line width-luminosity-mass
scaling relation of Kaspi \etal\ (2000).  The solid line represents
the \msigma\ relation from Tremaine \etal\ (2002).
\label{grupefig}}
\vskip +5mm

\noindent
(Ho et al.1997a,b), we see that the distribution of bulge properties
and nuclear emission-line widths are comparable for both Seyfert
types.  As shown in Table 1B of Ho et al. (2003), there is no
statistical difference between the Type 1 and Type 2 objects in terms
of Hubble type, absolute magnitude, and bulge luminosity, implying
identical distributions of \sigmastar\ between the two groups.  At the
same time, the distribution of NLR line widths (see Table 2B of Ho et
al. 2003) are also statistically equivalent.  The similarities in both
bulge and nuclear gas properties strongly suggest that Type 1 objects
will show the same distribution in \meansig, as that seen here for
Type 2 sources.

Second, we can directly test the hypothesis that \meansig\ is
invariant with respect to Seyfert type by using the study of Nelson \&
Whittle (1996), which provides \sigmastar\ and \sigmagas\ for 22 Type
1 and 44 Type 2 objects (these are the 66 objects with a tabulated
\oiii\ line width and stellar velocity dispersion).  Again, only FWHM
measurements are available.  We compare FWHM$_{\mathrm{[O \tiny
III]}}$ with \sigmastar\ for these objects in Figure 5.  The two
samples largely appear to overlap.  For Type 1 objects, \meansig$=
1.00 \pm 0.46$ while for Type 2 objects \meansig $=1.22 \pm 0.78$.  We
use the K-S test to compare the distribution of \sigmagas/\sigmastar\
for each group and find that there is no statistically significant
difference between the two; the probability of rejecting the null
hypothesis that the two groups are drawn from the same parent
population is $P_{\rm KS} = 13 \%$.  When we remove the four obvious
outliers, which are the powerful linear radio sources discussed by
Whittle (1992b), the similarity between the two samples becomes even
more apparent ($P_{\rm KS} = 16 \%$).

\epsfig{file=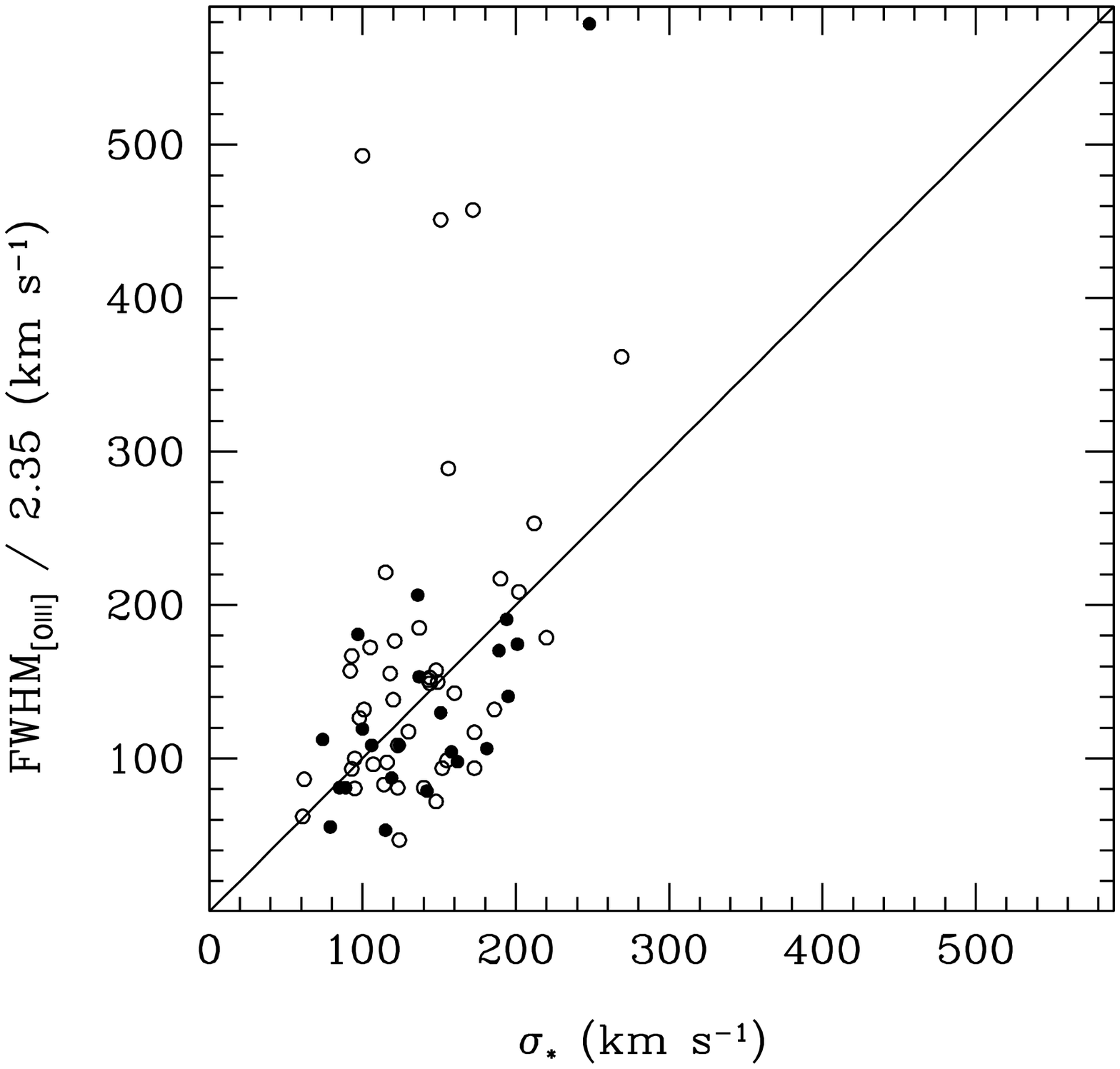,width=0.5\textwidth,keepaspectratio=true}
\vskip -0.1mm 
\figcaption[]{ 
\sigmagas\ = FWHM$_{\mathrm{[O \tiny III]}}$/2.35 plotted against \sigmastar\
for the Nelson \& Whittle (1996) sample.  Open symbols are Type 2
objects; filled symbols are Type 1 objects.  The solid line denotes
\sigmagas\ = \sigmastar. The extreme outliers are the powerful linear
radio sources discussed by Whittle (1992b).
\label{nelsfig}}
\vskip +5mm

We can also ask whether \delsig\ is correlated with Eddington ratio in
Type 1 objects using the Nelson \& Whittle (1996) sample.  In Figure
3{\it b}\ we have overplotted the Type 1 objects from the Nelson \&
Whittle sample with measured \sigmastar, $L_{\mathrm{[O \tiny III]}}$,
and FWHM$_{\mathrm{[O \tiny III]}}$.  Seven of the velocity
dispersions (NGC 3227, 3516, 4051, 4151, 4593, 5548, and Mrk 590) were
updated from the more recent work of Nelson \etal\ (2004).  We
corrected the \oiii\ luminosity to our assumed cosmology.  The
correlation is not significant for the Type 1 objects ($\rho = 0.27$,
$P_{\rm null} = 0.21$), possibly because of the small number of objects
considered, but they seem to trace roughly the trend defined by the Type 2 
objects.

\subsection{Aperture Effects}

Because the SDSS aperture includes the central $\sim 5$~kpc of the galaxy
a serious concern in the use of SDSS velocity dispersions is that
rather than measuring the bulge velocity dispersion at a well-defined
radius, a distance-dependent fraction of disk light may be included in the
measurements.  Both
nuclear gas and stellar velocity dispersions may vary with radius.
Therefore, we must examine the extent to which non-nuclear light
contamination matters. Below we examine both the luminosity and
kinematic distributions of stars and gas within the SDSS aperture to
argue that our measurements are meaningful despite the large aperture.
Rotational broadening is further discussed in the Appendix.

In the case of the gas, we expect the vast majority of the emission to
come from the NLR at sub-kpc scales.  After all, our sample was
selected to show unambiguous AGN-like line ratios.  Star-forming
regions are the strongest contaminant, \footnote{We note that of the
three tracers considered here, [O {\tiny II}] is most susceptible to
contamination by star formation.  Only in low-metallicity H {\tiny II}
regions, which are unlikely in the massive hosts of AGNs, is [O {\tiny
III}] a major coolant, while [S {\tiny II}] is strong in H {\tiny II}
regions only in the presence of shocks (e.g.,~Ho et al. 1997c; Kewley
\etal\ 2001).} but if emission from star formation were important,
then the spectra would look like those of \hii\ regions or transition
objects rather than bona fide AGNs. We can use the relation between
NLR radius ($R_{\mathrm{NLR}}$) and AGN luminosity to estimate the
typical radius dominating the gas emission.  Schmitt \etal\ (2003),
using a sample of 60 Seyferts of both types imaged with the {\it
Hubble Space Telescope}, find log $R_{\mathrm{NLR}}$ = (0.33 $\pm$
0.04) $\times$ log \loiii\ $-$ (10.78 $\pm$ 1.80), with
$R_{\mathrm{NLR}}$ in pc and \loiii\ in erg~s$^{-1}$.  The resulting
range in physical size for our objects is $\sim 0.25 - 1.2$ kpc.  At
$z = 0.1$, the corresponding angular size ranges from 0\farcs14 to
0\farcs67, well within the SDSS aperture.  Thus, we expect the radial
variation in \sigmagas\ to be negligible.

We now turn to the luminosity distribution of starlight within the
aperture.  Unlike the gas, the stars are distributed throughout the
aperture, and the velocity dispersion we measure may differ
substantially from \sigmastar\ at $R_{\rm e}$.  We estimate the sizes
of the bulges in our sample, and then infer the probable contamination
to \sigmastar\ from disk light.  For the median stellar velocity
dispersion of $\sim 150$ \kms\ in our sample, the fundamental plane
relation from Bernardi \etal\ (2003b) predicts $R_{\rm e} \approx 4.5$
kpc.  Alternatively, for a median galaxy mass of $\sim 5\times
10^{10}$ \msun, Kauffmann \etal\ (2003b) find a median $R_{\rm e}
\approx 3.3$ kpc.  If the bulges are characterized by $R_{\rm e}
\approx 4$~kpc then the majority of the starlight in the $\sim 5.4$
kpc aperture originates in the bulge, and we are not probing stars
much beyond an effective radius.

Even if the majority of starlight comes from the bulge, we still must
consider the radial dependence of \sigmastar\ with radius. J\o
rgensen, Franx, \& Kj\ae gaard (1995) use 51 E/S0 galaxies available
in the literature to derive an empirical aperture correction.  For
E/S0 galaxies, the integrated \sigmastar\ decreases outward as a
function of radius, falling as a power law with slope $-0.04$,
relative to an aperture with a radius of $R_{\rm e}/8$.  On the other
hand, Pizzella \etal\ (2004) find that the radial distribution in
velocity dispersion is flat or slightly increasing outward in a sample
of spiral (Sb-Sc) galaxies (see their Fig. 5).  In early-type galaxies
a large aperture will tend to underestimate \sigmastar\ for the bulge,
while for late-type galaxies there may be a slight tendency in the
opposite sense.  Since our sample consists of early-type galaxies, on
average the aperture variation most likely causes us to underestimate
\sigmastar.  Bernardi \etal\ (2003a) find a mean aperture correction
of $\sim 7 \%$ for a sample of early-type SDSS galaxies based on the
empirical function from J\o rgensen \etal\ (1995).  In any case, the
radial distribution of \sigmastar\ is quite flat.  As a check, we can
look for possible trends in \meansig\ as a function of redshift.
Since galaxy contamination, on average, should increase as a function
of redshift, we might expect \meansig\ to increase at higher redshift,
if we assume \sigmagas\ = \sigmastar\ intrinsically.  We show
\meansig\ in three redshift bins ($z < 0.1, 0.1 \leq z \leq 0.2$, and
$z > 0.2$) in Table 1.  We see no significant trend.

\section{Concluding Remarks}

Using a large sample of Type 2 AGNs selected from SDSS, we have
systematically investigated the relation between stellar velocity
dispersions (\sigmastar) and the kinematics of the emission-line gas
(\sigmagas) as traced by optical forbidden lines (\oii\ $\lambda3727$,
\oiii\ $\lambda5007$, and \sii\ $\lambda\lambda 6716, 6731$).  We show
that the widths of the low-ionization lines are gravitationally
dominated, and may be used as a proxy for \sigmastar.  We find that if
the blue asymmetric wing of the \oiii\ line is removed, the width of
the core component can also be used to estimate \sigmastar.  The
ionized gas in the central regions of active galaxies is evidently in
approximate pressure support with the gravitational potential of the
bulge stars of the host galaxy, consistent with the situation in
inactive early-type galaxies (e.g.,~Bertola \etal\ 1995; Cinzano
\etal\ 1999).  The scatter in each of the \sigmagas-\sigmastar\
relations is large, however, and we advocate their use only in
statistical studies.

While not the primary goal of this study, our results offer some
insights into the interaction between the AGN and the gas of the
circumnuclear region.  The nearly ubiquitous asymmetric blue wing of
the \oiii\ line has long been recognized as indicative of radial
motions in the NLR, presumably associated with an outflow, In this
study, we have shown that this outflowing component is principally
responsible for imparting supervirial motions to the gas.  That lines
with lower critical density generally lack this component suggests
that it probably originates from a more compact region closer to the
center.  But what is the main physical driver of this component?  We
have shown that the most important variable appears to be the
Eddington ratio.  This finding may offer a useful outer boundary
condition for constraining disk-wind models of the BLR (e.g.,~Murray
\& Chiang 1997; Proga, Stone, \& Drew 1998; Proga, Stone, \& Kallman
2000).

\acknowledgements 
We thank an anonymous referee for comments that
helped to clarify the manuscript. We are grateful for useful
conversations with A.~J.~Barth, P. Martini, R. Narayan, A. Pizzella,
and D.~Proga.  J.~E.~G. is funded by a National Science Foundation
Graduate Research fellowship.  L.~C.~H. acknowledges support by the
Carnegie Institution of Washington and by NASA grants from the Space
Telescope Science Institute (operated by AURA, Inc., under NASA
contract NAS5-26555).  We are grateful to L. Hao for making available
her principal component analysis software, and to the entire SDSS
collaboration for providing the extraordinary database and processing
tools that made this work possible.

\newpage
\appendix
\section{Rotational Broadening}

Rotation must play a role in the nuclear kinematics.  In particular,
it is not unreasonable to expect some fraction of the nuclear gas to reside 
in a disk-like component.  Whittle (1992b) concludes that rotation
contributes at least partially to the \oiii\ line width in order to
reproduce the observed relations between rotational velocity,
inclination, and line width.  Vega Beltr\'{a}n \etal\ (2001) find that the
ionized gas in a sample of 20 late-type disk galaxies is dominated by
rotation.  Without rotation velocities or inclination angles for our 
sample, it is difficult to directly examine the rotational
contribution.  Only the double-peaked emission lines provide evidence
that rotation can dominate the kinematics, though they are only
readily apparent in $\sim 1 \%$ of the cases.  However, we worry that 
rotational broadening may artificially increase our observed line widths.  
The same concern applies to the stellar velocity dispersions, since the 
SDSS aperture is sufficiently large that some disk light must contaminate 
the bulge \sigmastar.

We can estimate the contribution of rotational broadening to the
observed scatter from both stellar and gaseous components by
constructing a toy galaxy model with the average properties of our sample.
In particular, we need a model for both the luminosity and
radial velocity distributions in the disk.  The
rotational width is then simply

$$ \sigma_{\mathrm{rot}}^2 = \int (v - {\bar v})^2~I~\mathrm{d}v / 
\int I~\mathrm{d}v, $$

\noindent
where $I$ and $v$ are both functions of radius, and ${\bar v}$ is the
luminosity-weighted mean velocity.  For the intensity, we use an
exponential distribution, $I = I_0$ exp ($-R/R_\mathrm{d}$), with an
exponential scale length of $R_\mathrm{d} \approx 3$~kpc (Neistein
\etal\ 1999).  Since the gas emission is dominated by the NLR, the gas
emission radii range from 0.25--1.2 kpc, while the stellar disk
extends to a projected aperture radius of 2.7 kpc at $z=0.1$.

The maximum circular rotational velocity, \vc, may be estimated using the
Tully-Fisher relation.  In order to account for the possible
contamination of the host luminosity from the AGN, we use two
different estimates of the absolute magnitude of the underlying host
galaxy.  The first uses the measured, K-corrected $M_z$ values from
Brinchmann et al.  (2004a), corrected to our assumed cosmology.  This
yields an average $M_z = -21.98$ mag.  One may worry, however, that
AGN contamination is present even in Type 2 objects.  We therefore
indirectly estimate $M_z$ in the following way.  From the relation
between concentration index and $(M/L)_z$ shown in Figure 13 of
Kauffmann \etal\ (2003a) we can estimate the probable range in
$(M/L)_z$.  For our estimated mean concentration index of $\sim 2.65$,
we adopt a range in logarithmic mass-to-light ratios of $(M/L)_z$ =
$-0.1$ to $0.2$.  If we assume that the typical host galaxy has a mass
of $\sim 10^{10.7}$ \msun, we find a range in absolute
magnitude\footnote{To convert luminosity to absolute magnitude we
assume, following Kauffmann \etal\ (2003a), that the $z$ absolute
magnitude of the Sun is $M_{z~\sun}$ = 4.51 mag at $z$=0.1.} of
$-21.74 < M_z < -22.49$.  We are encouraged that the two methods are
consistent.  In what follows, we adopt $M_z = -22$ mag.  Giovanelli
\etal\ (1997) give the following form for the $I$-band Tully-Fisher
relation: $M_I = -21 - 7.52 x$, where $x =$ log (\vc/2) $-$ 2.5.  We
estimate $M_I$ using the color correction for S0 galaxies from
Fukugita, Shimasaku, \& Ichikawa (1995), $z - I_{\rm C} = 0.21$ mag.
Our resulting average \vc\ is 230 \kms.  While we argue that our
sample is dominated by early-type galaxies, we note that recent
studies of the Tully-Fisher relation in S0 galaxies have found
remarkably small offsets from the late-type population (Neistein
\etal\ 1999; Hinz, Rieke, \& Caldwell 2003).

As shown by Sofue \etal\ (1999), the rotation curves of active
galaxies are comparable to those of quiescent galaxies, and so we use
their measured rotation curves of barred and unbarred Sb and Sc
galaxies as templates for our model.  As a worst-case scenario, we
only consider edge-on disks with \vc\ = 230~\kms.  The radius at which
the rotation curve flattens is allowed to vary between 0.1 and 1 kpc,
but the rising portion of the rotation curve need not be monotonic; in
one of our models the velocity peaks at $\sim 0.3$~kpc, and then dips
again before flattening at $\sim 2.5$~kpc.  In the case of the gas,
assuming emission radii between 0.25 and 1.2 kpc,
$\sigma_{\mathrm{rot}}$ ranges from 8 to 74~\kms.  The measured
$\sigma$ is simply the quadrature sum of the true \sigmagas\ and
$\sigma_{\mathrm{rot}}$, so a measured \sigmagas\ $\approx 150$~\kms\
may be overestimated by up to $15 \%$.  To estimate
$\sigma_{\mathrm{rot}}$ in stars, we assume the same family of
rotation curves but fix the emission radius at 2.7~kpc, for a maximum
error of $\sim 12 \%$ in \sigmastar.  If we allow each component to
independently assume its maximum error, and examine the case where
\sigmagas\ = \sigmastar\ = 150 \kms, we infer that the observed
\sigmagas/\sigmastar\ may vary from 0.9 to 1.1.  We therefore conclude
that rotational broadening contributes to, but cannot fully account
for, the observed scatter.


\clearpage

\begin{thebibliography}{}

\bibitem[]{}Abazajian, K., et al. 2004, \aj, 128, 502

\bibitem[]{}Adams, T.~F., \& Weedman, D.~W. 1975, \apj, 199, 19

\bibitem[]{}Akritas, M.~G., \& Siebert, J. 1996, \mnras, 278, 919

\bibitem[]{}Antonucci, R.~R.~J., \& Miller, J.~S.\ 1985, \apj, 297, 621 

\bibitem[]{}Baldwin, J.~A., Phillips, M.~M., \& Terlevich, R. 1981,
\pasp, 93, 5

\bibitem[]{}Barth, A.~J., Greene, J.~E., \& Ho, L.~C. 2005, \apjl, 619, L151 

\bibitem[]{}Becker, R.~H., White, R.~L., \& Helfand, D.~J. 1995, \apj,
450, 559

\bibitem[]{}Bernardi, M., et~al. 2003a, \aj, 125, 1817

\bibitem[]{}------. 2003b, \aj, 125, 1866

\bibitem[]{}Bertola, F., Cinzano, P., Corsini, E.~M., Rix, H.-W., \&
	Zeilinger, W.~W. 1995, \apjl, 448, L13

\bibitem[]{}Boroson, T.~A. 2002, \apj, 565, 78

\bibitem[]{}Brinchmann, J., Charlot, S., Heckman, T.~M., Kauffmann,
	     G., Tremonti, C., \& White, S.~D.~M. 2004a, (astro-ph/0406220)

\bibitem[]{}Brinchmann, J., Charlot, S., White, S.~D.~M., Tremonti,
	     C., Kauffmann, G., Heckman, T., \& Brinkmann, J. 2004b, 
	     \mnras, 351, 1151

\bibitem[]{}Caldwell, N., Kirshner, R.~P., \& Richstone, D.~O. 1986,
	\apj, 305, 136

\bibitem[]{}Cinzano, P., Rix, H.-W., Sarzi, M., Corsini, E.~M.,
Zeilinger, W.~W., \& Bertola, F. 1999, \mnras, 307, 433

\bibitem[]{}Cinzano, P., \& van der Marel, R.~P. 1994, \mnras, 270, 325

\bibitem[]{}de Bruyn, A.~G., \& Wilson, A.~S. 1978, \aap, 64, 433

\bibitem[]{}De Robertis, M.~M., \& Osterbrock, D.~E. 1984, \apj, 286, 171

\bibitem[]{}------. 1986, \apj, 301, 98

\bibitem[]{}Ferrarese, L., \& Merritt, D. 2000, \apj, 539, L9

\bibitem[]{}Ferrarese, L., Pogge, R.~W., Peterson, B.~M., Merritt, D.,
	Wandel, A., \& Joseph, C.~L. 2001, \apj, 555, L79

\bibitem[]{}Filippenko, A.~V. 1985, \apj, 289, 475

\bibitem[]{}Filippenko, A.~V., \& Halpern, J.~P. 1984, \apj, 285, 458

\bibitem[]{}Fillmore, J.~A., Boroson, T.~A., \& Dressler, A. 1986,
	\apj, 302, 208

\bibitem[]{}Fisher, D. 1997, \aj, 113, 950

\bibitem[]{}Fukugita, M., Ichikawa, T., Gunn, J.~E., Doi, M., Shimasaku, K.,
        \& Schneider, D.~P. 1996, \aj, 111, 1748

\bibitem[]{}Fukugita, M., Shimasaku, K., \& Ichikawa, T. 
1995, \pasp, 107, 945 

\bibitem[]{}Gebhardt, K., et~al. 2000a, \apjl, 539, L13

\bibitem[]{}------. 2000b, \apjl, 543, L5

\bibitem[]{}Giovanelli, R., Haynes, M.~P., Herter, T., Vogt, N.~P.,
da Costa, L.~N., Freudling, W., Salzer, J.~J., \& Wegner, G. 1997,
\aj, 113, 53 

\bibitem[]{}Greene, J.~E., \& Ho, L.~C. 2004, \apj, 610, 722

\bibitem[]{}Grupe, D., \& Mathur, S.\ 2004, \apjl, 606, L41

\bibitem[]{}Grupe, D., Wills, B.~J., Leighly, K.~M., \& Meusinger,
	H. 2004, 127,156

\bibitem[]{}Gunn, J. E., \etal\ 1998, \aj, 116, 3040

\bibitem[]{}Hao,~L., \etal\ 2005, \aj, in press (astro-ph/0501059) 

\bibitem[]{}Heckman, T.~M. 1980, \aap, 87, 152

\bibitem[]{}Heckman, T.~M., Kauffmann, G., Brinchmann, J., Charlot,
S., Tremonti, C., \& White, S.~D.~M. 2004, \apj, 613, 109

\bibitem[]{}Heckman, T.~M., Miley, G.~K., van Breugel, W.~J.~M., 
	\& Butcher, H.~R. 1981, \apj, 247, 403

\bibitem[]{}Hinz, J.~L., Rieke, G.~H., \& Caldwell, N. 2003, \aj, 126, 2622 

\bibitem[]{}
Ho, L.~C. 2004, in Carnegie Observatories Astrophysics Series, Vol. 1: 
Coevolution of Black Holes and Galaxies, ed. L. C. Ho (Cambridge: Cambridge
Univ. Press), 293

\bibitem[]{}Ho, L.~C., Filippenko, A.~V., \& Sargent, W.~L.~W. 1993, \apj, 417, 63

\bibitem[]{}------. 1997a, \apjs, 112, 315

\bibitem[]{}------. 1997b, \apj, 487, 568

\bibitem[]{}------. 1997c, \apj, 487, 579

\bibitem[]{}------. 2003, \apj, 583, 159

\bibitem[]{}Ho, L.~C., \& Peng, C.~Y. 2001, \apj, 555, 650

\bibitem[]{}Ho, L.~C., \& Ulvestad, J.~S. 2001, \apjs, 133, 77

\bibitem[]{}Ivezi\'c, Z., et~al. 2002, \aj, 124, 2364

\bibitem[]{}J\o rgensen, I., Franx, M., \& Kj\ae rgaard, P.\ 1995, \mnras, 276, 1341

\bibitem[]{}Kaspi, S., Smith, P.~S., Netzer, H., Maoz, D., Jannuzi,
	B.~T., \& Giveon, U. 2000, \apj, 533, 631

\bibitem[]{}Kauffmann, G., et~al. 2003a, \mnras, 341, 33

\bibitem[]{}------. 2003b, \mnras, 341, 54

\bibitem[]{}------. 2003c, \mnras, 346, 1055

\bibitem[]{} Kauffmann, G., White, S.~D.~M., Heckman, T.~M., M{\' e}nard, B.,
Brinchmann, J., Charlot, S., Tremonti, C., \& Brinkmann, J. 2004, \mnras, 353, 713

\bibitem[]{}Kent, S.~M. 1988, \aj, 96, 514

\bibitem[]{}Kewley, L.~J., Dopita, M.~A., Sutherland, R.~S., Heisler,
	C.~A., \& Trevena, J. 2001, \apj, 556, 121

\bibitem[]{}Kormendy, J., \& Westpfahl, D.~J. 1989, \apj, 338, 752

\bibitem[]{}Marconi, A., Risaliti, G., Gilli, R., Hunt, L.~K., 
	   Maiolino, R., \& Salvati, M. 2004, \mnras, 
	   351, 169

\bibitem[]{}Mathur, S., Kuraszkiewicz, J., \& Czerny, B.\ 2001, New
	     Astronomy, 6, 321

\bibitem[]{}McLure, R. J., \& Dunlop, J. S. 2004, MNRAS, 352, 1390

\bibitem[]{}Murray, N., \& Chiang, J. 1997, \apj, 474, 91

\bibitem[]{}Neistein, E., Maoz, D., Rix, H.-W., \& Tonry, J.~L.
1999, \aj, 117, 2666 

\bibitem[]{}Nelson, C.~H. 2000, \apjl, 544, L91

\bibitem[]{}Nelson, C.~H., Green, R.~F., Bower, G., Gebhardt, K.,
	\& Weistrop, D. 2004, \apj, 615, 652

\bibitem[]{}Nelson, C.~H., \& Whittle, M. 1996, \apj, 465, 96

\bibitem[]{}Onken, C.~A., Ferrarese, L., Merritt, D., Peterson, B.~M.,
	Pogge, R.~W., Vestergaard, M., \& Wandel, A. 2004, \apj, 615, 645 

\bibitem[]{}Osterbrock, D.~E. 1989, Astrophysics of Gaseous Nebulae
	and Active Galactic Nuclei (Mill Valley: University Science Books)

\bibitem[]{}Pelat, D., Alloin, D., \& Fosbury, R.~A.~E. 1981, \mnras, 195, 787
				   
\bibitem[]{}Peterson, B.~M., \etal\ 2004, \apj, 613, 682

\bibitem[]{}Pizzella, A., Corsini, E.~M., Vega 
	Beltr{\'a}n, J.~C., \& Bertola, F. 2004, \aap, 424, 447

\bibitem[]{}Pounds, K.~A., Done, C., \& Osborne, J.~P. 1995, \mnras,
	277, L5

\bibitem[]{}Proga, D., Stone, J.~M., \& Drew, J.~E. 1998, \mnras, 295, 595 

\bibitem[]{}Proga, D., Stone, J.~M., \& Kallman, T.~R. 2000, \apj, 543, 686 

\bibitem[]{}Rees, M.~J., Phinney, E.~S., Begelman, M.~C., \&
	Blandford, R.~D., 1982, \nat, 295, 17

\bibitem[]{} Schlegel, D.~J., Finkbeiner, D.~P., \& Davis, M. 1998, 
\apj, 500, 525 

\bibitem[]{}Schmitt, H.~R., Donley, J.~L., Antonucci, R.~R.~J.,
Hutchings, J.~B., Kinney, A.~L., \& Pringle, J.~E. 2003, \apj, 597, 768 

\bibitem[]{}Sheth, R. K., et al. 2003, \apj, 594, 225

\bibitem[]{}Shields, G.~A., Gebhardt, K., Salviander, S., Wills, 
	B.~J., Xie, B., Brotherton, M.~S., 
	Yuan, J., \& Dietrich, M.\ 2003, \apj, 583, 124

\bibitem[]{}Shimasaku, K., et~al. 2001, \aj, 122, 1238

\bibitem[]{}Smith, E.~P., Heckman, T.~M., \&  Illingworth, G.~D. 1990,
	\apj, 356, 399

\bibitem[]{}Smith, J.~A., \etal\ 2002, \aj, 123, 2121

\bibitem[]{}Sofue, Y.~,\& Rubin, V. 2001, \araa, 39, 137 

\bibitem[]{}Sofue, Y., Tutui, Y., Honma, M., Tomita, A., Takamiya, T., Koda, J., \& Takeda, Y. 1999, \apj, 523, 136 

\bibitem[]{}Stoughton, C., \etal\ 2002, \aj, 123, 485

\bibitem[]{}Strateva, I., et~al. 2001, \aj, 122, 1861

\bibitem[]{}Strauss, M.~A., \etal\ 2002, \aj, 124, 1810

\bibitem[]{}Tadhunter, C.~N., \& Tsvetanov, Z. 1989, \nat, 341, 422 

\bibitem[]{}Tremaine, S., et al. 2002, \apj, 574, 740

\bibitem[]{}Ulvestad, J.~S., \& Wilson, A.~S. 1984, \apj, 278, 544

\bibitem[]{}Vega Beltr{\'a}n, J.~C., Pizzella, A., Corsini, E.~M.,
	Funes, J.~G., Zeilinger, W.~W., Beckman, J.~E., \& Bertola, F.
	2001, \aap, 374, 394

\bibitem[]{}Veilleux, S., \& Osterbrock, D.~E.\ 1987, \apjs, 63, 295

\bibitem[]{}Wang, T., \& Lu, Y.\ 2001, \aap, 377, 52 

\bibitem[]{}Whittle, M. 1985a, \mnras, 213, 1

\bibitem[]{}------. 1985b, \mnras, 213, 33

\bibitem[]{}------. 1992a, \apj, 387, 109

\bibitem[]{}------. 1992b, \apj, 387, 121

\bibitem[]{}Wilson, A.~S., \& Heckman, T.~M. 1985, in Astrophysics of
	Active Galaxies and Quasi-Stellar Objects, ed. J.~S.~Miller
	(Mill Valley: University Science Books), 39

\bibitem[]{}Wilson, A.~S., \& Willis, A.~G. 1980, \apj, 240, 429

\bibitem[]{}Yee, H.~K.~C. 1980, \apj, 241, 894 

\bibitem[]{}Yee, H.~K.~C., \& Oke, J.~B. 1978, \apj, 226, 753

\bibitem[]{}York, D.~G., et~al. 2000, \aj, 120, 1579

\bibitem[]{}Yu, Q., \& Tremaine, S. 2002, \mnras, 335, 965

\bibitem[]{}Zakamska, N., \etal\ 2003, \aj, 126, 2125

\end{thebibliography}
\end{document}